\documentclass[epjCONF,columns]{svjour} % for 2 columns format
\usepackage{graphics}
\usepackage[varg]{txfonts} % Times fonts
\usepackage[latin1]{inputenc}
\session-title{Hadron Collider Physics Symposium 2011}
\usepackage[amssymb,thinspace]{SIunits}
\usepackage{units}
\usepackage{xspace}
\usepackage{fixltx2e}
\newcommand{\pT}{\ensuremath{p_T}\xspace}
\newcommand{\ET}{\ensuremath{E_T}\xspace}
\newcommand{\TeV}{\tera\electronvolt\xspace}
\newcommand{\GeV}{\giga\electronvolt\xspace}

\newcommand{\fb}{\femto\barn\xspace}
\newcommand{\fbi}{\ensuremath{\fb^{-1}}\xspace}

\newcommand{\MET}{\ensuremath{\ET^{\textnormal{miss}}}\xspace}
\newcommand{\HT}{\ensuremath{H_T}\xspace}
\newcommand{\cls}  {\ensuremath{\mathrm{CL_S}}\xspace}
\newcommand{\CMSSM}  {CMSSM\xspace}
\newcommand{\ttbar}  {top-antitop\xspace}
\newcommand{\pTll}{\ensuremath{\pT(ll)}\xspace}
\begin{document}
\title{Search for new physics in events with opposite-sign dileptons and missing transverse energy with the CMS experiment}
\author{Daniel Sprenger (on behalf of the CMS Collaboration)}
\institute{Physics Institute 1B of RWTH Aachen University}
\abstract{
The results of a search for new physics in events with two opposite-sign isolated electrons or muons, hadronic activity, and missing transverse energy in the final state are presented. The results are based on analysis of a data sample with a corresponding integrated luminosity of 0.98~\fbi produced in pp collisions at a center-of-mass energy of 7~\TeV collected by the CMS experiment at the LHC. No evidence for an event yield beyond Standard-Model expectations is found, and constraints on supersymmetric models are deduced from these observations.
} %end of abstract
\maketitle
\section{Introduction}
\label{sec:introduction}

A search for physics beyond the standard model in final states with opposite-sign isolated lepton pairs accompanied by hadronic jets and missing transverse energy is presented~\cite{PAS}. The search is based on LHC data recorded with the CMS experiment corresponding to an integrated luminosity of 0.98~\fbi.

Two complementary search strategies are performed. The first search probes models with heavy, colored objects which decay to final states including invisible particles, leading to very large hadronic activity and missing transverse energy. The second search probes models with a specific dilepton production mechanism, which leads to a characteristic kinematic edge in the dilepton-mass distribution.

No specific model has been used to optimise this search. However, Supersymmetry (SUSY) models and benchmark points have been chosen to illustrate the sensitivity of the search.

\section{The CMS detector}
\label{sec:cms}

The central feature of the Compact Muon Solenoid (CMS) apparatus is a superconducting solenoid, of 6~m internal diameter, providing a field of 3.8~T. Within the field volume are the silicon pixel and strip tracker, the crystal electromagnetic calorimeter (ECAL) and the brass/scintillator hadron calorimeter (HCAL). Muons are measured in gas-ionization detectors embedded in the steel return yoke. In addition to the barrel and endcap detectors, CMS has extensive forward calorimetry. A much more detailed description of CMS can be found elsewhere~\cite{CMS}.

\section{Counting experiment}
\label{sec:counting}

A cut-and-count analysis is used as a generic approach to search for physics beyond the standard model.

\subsection{Event selectiont}
\label{sec:counting:eventselection}

For basic event selection, a dilepton trigger for light leptons (electrons and muons) is used. Two isolated leptons with opposite charge sign are demanded. One of these leptons has to have a transverse momentum of at least 20~\GeV, the other lepton of at least 10~\GeV. To suppress low resonances and Z boson background, the invariant mass of the two leptons is required to be larger than 12~\GeV and outside the Z boson mass window (76~\GeV to 106~\GeV).

% (``high \MET region'')
% (``high \HT region'')
Two signal regions are defined. The first region uses a very tight cut on missing transverse energy, \MET, of 275~\GeV and a moderate cut on the summed up momenta of all jets, \HT, of 300~\GeV. The second signal region tightens the \HT cut to 600~\GeV while loosening the \MET cut to 200~\GeV.

\subsection{Background estimation}
\label{sec:counting:background}

Two data-driven methods~\cite{OldPaper} are used to predict standard model background in the signal regions.

The first method (``ABCD prediction'') measures the background yield dependence on \HT and $y=\MET / \sqrt{\HT}$ in two control regions. These variables are found to be uncorrelated. Therefore, using the determined dependencies, the background yield can be extrapolated from a control region into the signal regions.

The second method (``\pTll prediction'') exploits the fact that the \MET distribution in \ttbar events can be modeled using the \pT distribution of the charged leptons in these events~\cite{pTll}. \MET is described as the sum of the neutrino transverse momenta in these events, which follows the same distribution as the charged leptons. Using correction factors for detector effects, W polarisation and the applied \MET cut, the amount of \ttbar background in the signal regions can be extrapolated with this method.

\subsection{Results}
\label{sec:counting:results}

The observed yields and prediction results of both background estimation methods are shown in Table~\ref{tab:yields} for both signal regions.

\begin{table*}[hbt]
\begin{center}
\caption{Observed and predicted yields of the counting experiment in the two signal regions. The background yield, $N_{bkg}$ is the error-weighted average of the two data-driven background prediction methods. The non-standard-model yield upper limit (UL) is a \cls 95\%-confidence-level upper limit.}
\label{tab:yields}       % Give a unique label
% For LaTeX tables use
\begin{tabular}{lcc}
\hline\noalign{\smallskip}
% & high \MET signal region & high \HT signal region  \\
 & ($\HT > 300~\GeV$, $\MET > 275~\GeV$) & ($\HT > 600~\GeV$, $\MET > 200~\GeV$) \\
\noalign{\smallskip}\hline\noalign{\smallskip}
observed yield & 8 & 4\\\hline
MC prediction & $7.3 \pm 2.2$ & $7.1 \pm 2.2$\\
ABCD prediction & $4.0 \pm 1.0 \textnormal{(stat)} \pm 0.8 \textnormal{(sys)}$ & $4.5 \pm 1.6 \textnormal{(stat)} \pm 0.9 \textnormal{(sys)}$\\
\pTll prediction & $14.3 \pm 6.3 \textnormal{(stat)} \pm 5.3 \textnormal{(sys)}$ & $10.1 \pm 4.2 \textnormal{(stat)} \pm 3.5 \textnormal{(sys)}$\\
$N_{bkg}$ & $4.2 \pm 1.3$ & $5.1 \pm 1.7$\\\hline
non-SM yield UL & 10 & 5.3\\\hline
LM1 & $49 \pm 11$ & $38 \pm 12$\\
LM3 & $18 \pm 5$ & $19 \pm 6$\\
LM6 & $8.1 \pm 1.0$ & $7.4 \pm 1.2$\\
\noalign{\smallskip}\hline
\end{tabular}
\end{center}
\end{table*}

%\begin{figure}
%\resizebox{0.75\columnwidth}{!}{%
%\resizebox{0.9\columnwidth}{!}{%
%  \includegraphics{figures/countingYieldsPAS} }
%\caption{Please write your figure caption here.}
%\label{fig:counting:yields}
%\end{figure}

\begin{figure}
%\resizebox{0.75\columnwidth}{!}{%
\resizebox{0.9\columnwidth}{!}{%
  \includegraphics{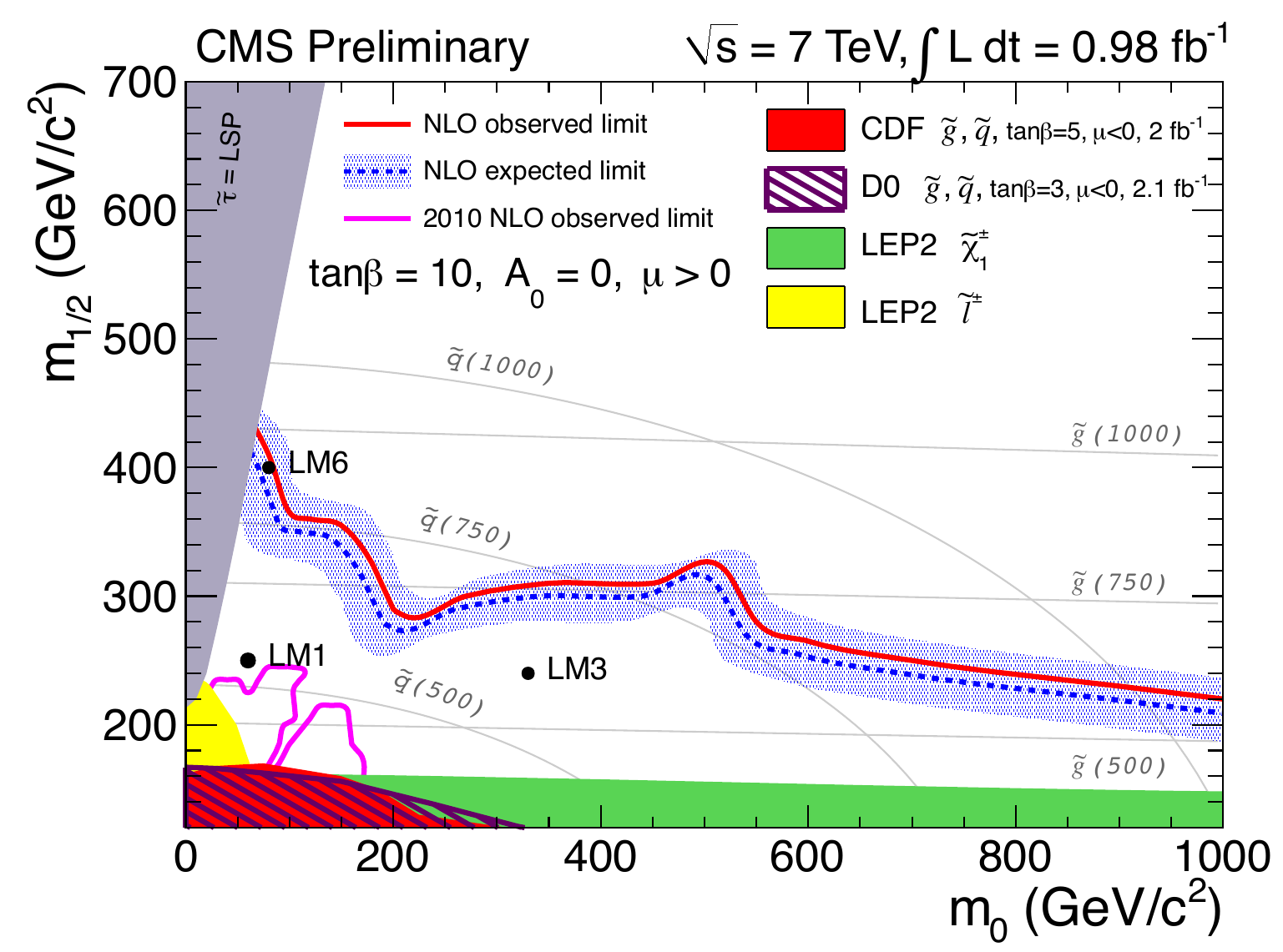} }
\caption{Contour of the observed 95\%-confidence-level exclusion and the expected limit for the counting experiment in the \CMSSM $m_0-m_{1/2}$ plane. The area below the curve is excluded.}
\label{fig:counting:limit}
\end{figure}

No evidence for a signal is found, and upper limits on non-standard-model yields in the signal regions are derived (Tab.~\ref{tab:yields}). A hybrid frequentist-bayesian \cls method~\cite{CLS} is used for this. In a further step, limits on the parameters of the \CMSSM~\cite{CMSSM1,CMSSM2} are derived (Fig.~\ref{fig:counting:limit}).

The CMS benchmark points~\cite{PhysTDR} LM1 and LM6 are in the $\tan{\beta} = 10$ plane, LM3 is located at $\tan{\beta} = 20$. For all three benchmark points is $A_0 = 0$ and $\mu > 0$. LM1, LM3 and LM6 are excluded.

\section{Search for kinematic edge}
\label{sec:edge}

Dilepton events are selected using the same basic selection as for the counting experiment (Sec.~\ref{sec:counting:eventselection}), and the invariant mass distribution of the lepton pair is investigated and searched for a kinematic edge. An edge in the invariant mass distribution is characteristic in SUSY models for decays of Neutralinos into pairs of flavour-correlated leptons e.g. $\tilde \chi^0_2 \rightarrow \tilde l l \rightarrow \tilde \chi^0_1 l^+ l^-$.

Due to the flavour correlation in the signal process, non-standard-model production of uncorrelated lepton pairs is considered a background to this search.

\subsection{Method}
\label{sec:edge:method}

An unbinned maximum-likelihood fit is performed to the dilepton invariant mass distribution of same-flavour lepton pair events in order to estimate the number of signal events on top of the standard-model background. The fit consists of three components.
Flavour-uncorrelated background e.g. by \ttbar production is the main background. It is modeled by the product of a power function and a falling exponential and estimated using different-flavour events.  Z-boson background is modeled using a Breit-Wigner function convolved with a Gaussian. The third component represents the non-standard-model signal and is modeled by a triangular shape with an endpoint at $m_{ll} = 78~\GeV$. This corresponds to the mass edge of CMS benchmark point LM1. A model for separate invariant mass resolutions in $ee$ and $\mu\mu$ events of $\sigma_{ee} = 2 \pm 1~\GeV$ and $\sigma_{\mu\mu} = 1 \pm 0.5~\GeV$ is included in this component.

The fit is performed simultaneously to $ee$, $e\mu$ and $\mu\mu$ dilepton events.

\subsection{Control and signal region}
\label{sec:counting:controlregion}

The method is tested in a control region that is defined by $100 < \HT < 300~\GeV$ and $\MET > 100~\GeV$. No signal is expected in this region. %Figure~\ref{fig:edge:control} shows the fit and its components to the data in the control region.
After performing the fit, a signal contribution of $10.7 \pm 15.4$ is obtained, which is compatible with the background-only hypothesis.

%\begin{figure*}[ht!]
%%\resizebox{0.75\columnwidth}{!}{%
%\resizebox{0.95\columnwidth}{!}{%
%  \includegraphics{figures/kinEdgeControlRegionSFOS} }
%\resizebox{0.95\columnwidth}{!}{%
%  \includegraphics{figures/kinEdgeControlRegionOFOS} }
%\caption{Results of the fit to the invariant mass distribution of same-flavour events (left) and different-flavour events (right) in the control region. The extracted number of signal, different-flavour- and Z background ($n_S$, $n_B$ and $n_Z$) is also shown.}
%\label{fig:edge:control}
%\end{figure*}

%\begin{figure}
%%\resizebox{0.75\columnwidth}{!}{%
%\resizebox{0.8\columnwidth}{!}{%
%  \includegraphics{figures/kinEdgeControlRegionOFOS} }
%\caption{Please write your figure caption here.}
%\label{fig:edge:controlofos}
%\end{figure}

%\subsection{Signal region}
%\label{sec:counting:signalregion}

\begin{figure*}[ht]
%\resizebox{0.75\columnwidth}{!}{%
\resizebox{0.99\textwidth}{!}{%
  \includegraphics{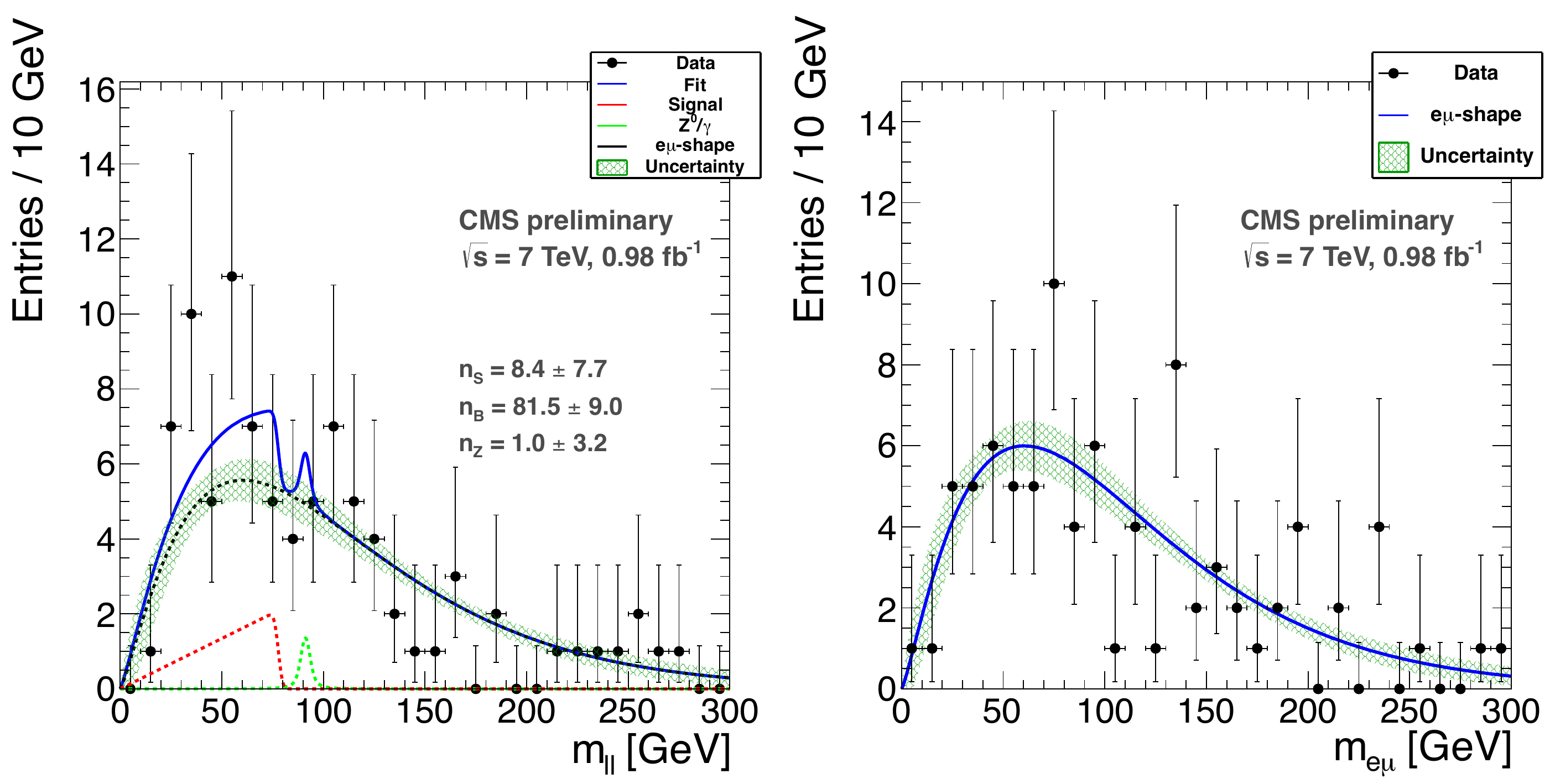} }
%\resizebox{0.99\columnwidth}{!}{%
%  \includegraphics{figures/kinEdgeSignalRegionOFOS} }
\caption{Results of the fit to the invariant mass distribution of $ee$ and $\mu\mu$ events (left) and $e\mu$ events (right) in the signal region. The three fit components of the same-flavour fit together with the extracted number of signal, different-flavour- and Z background ($n_S$, $n_B$ and $n_Z$) are displayed.}
\label{fig:edge:signal}
\end{figure*}

The signal region is defined by $\HT > 300~\GeV$ and $\MET > 100~\GeV$. Figure~\ref{fig:edge:signal} shows the fit and its components to the data in this region. A number of signal events of $8.4 \pm 7.7$ is extracted in this region. This is compatible with the background-only hypothesis; no evidence for a signal is found.

%\begin{figure}
%%\resizebox{0.75\columnwidth}{!}{%
%\resizebox{0.8\columnwidth}{!}{%
%  \includegraphics{figures/kinEdgeSignalRegionOFOS} }
%\caption{Please write your figure caption here.}
%\label{fig:edge:signalofos}
%\end{figure}

\subsection{Limit}
\label{sec:edge:limit}

After performing the fit for a mass edge with a cut-off at 78~\GeV, the range between 20~\GeV and
300~\GeV is scanned for other possible mass-edge cut-off points.
No evidence for a kinematic edge is found in this range.

\begin{figure}[h]
%\resizebox{0.75\columnwidth}{!}{%
\resizebox{0.9\columnwidth}{!}{%
  \includegraphics{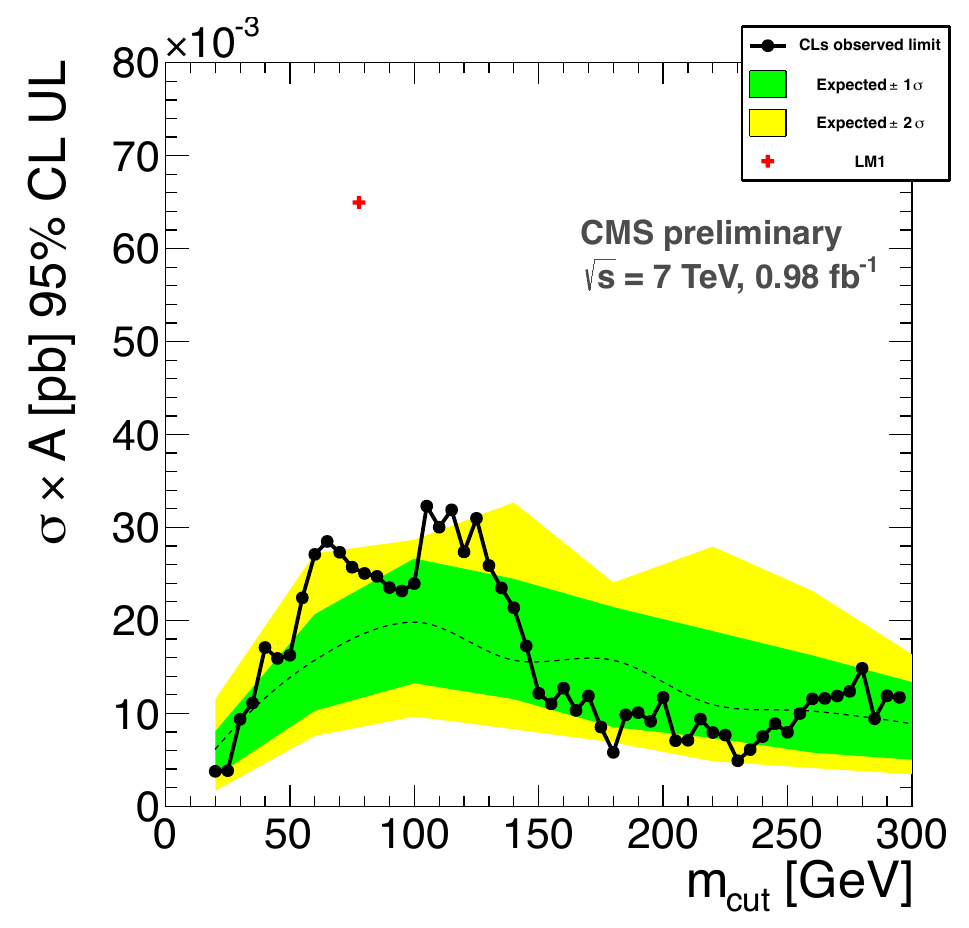} }
\caption{95\%-confidence-level observed and expected upper limit on cross-section times acceptance for the kinematic edge search. The dependence of the limit on the endpoint of the triangular mass edge, $m_{cut}$, is shown.}
\label{fig:edge:limit}
\end{figure}

Therefore limits on the production cross section of corresponding processes are set. The limits are estimated using a hybrid frequentist-bayesian \cls method~\cite{CLS} and are derived as a function of the cut-off parameter of the mass edge (Fig.~\ref{fig:edge:limit}).

\section{Summary}
\label{sec:summary}

Two complementary methods were presented to probe dilepton events with opposite charge sign for physics beyond the standard model. The first method focuses on models in which dilepton pairs are accompanied by significant hadronic activity and missing transverse energy. The other method focuses on models with a specific dilepton production mechanism, which leads to a characteristic edge in the invariant mass distribution of the lepton pair. No evidence for new physics is found and limits on non-standard-model physics are derived. 

Additional information related to detector efficiencies and response can be found in the CMS Physics Analysis Summary~\cite{PAS}. Using this it can be tested whether specific models of new physics are excluded by the presented results.

\end{document}